\documentstyle[12pt,
epsfig]{article}
\relax

\def\be{\begin{equation}}
\def\ee{\end{equation}}
\def\bs{\begin{subequations}}
\def\es{\end{subequations}}

\def\lx{\lambda}
\def\lh{\hat{l}}
\def\Rh{\hat{R}}

\def\Lx{\Lambda}
\def\dl{\delta_\lambda}
\def\dz{\delta_Z}
\def\dm{\delta_m}

\newcommand{\sx}{\sigma}

\newcommand{\rht}{\tilde{\rho}}

\def\be{\begin{equation}}
\def\ee{\end{equation}}
\def\bs{\begin{subequations}}
\def\es{\end{subequations}}
\newcommand{\een}{\end{subequations}}
\newcommand{\ben}{\begin{subequations}}
\newcommand{\beq}{\begin{eqalignno}}
\newcommand{\eeq}{\end{eqalignno}}

\def \lta {\mathrel{\vcenter
     {\hbox{$<$}\nointerlineskip\hbox{$\sim$}}}}
\def \gta {\mathrel{\vcenter
     {\hbox{$>$}\nointerlineskip\hbox{$\sim$}}}}

\def\mpl{M_{\rm Pl}}

\newcommand\fverb{\setbox\pippobox=\hbox\bgroup\verb}
\newcommand\fverbdo{\egroup\medskip\noindent%
                        \fbox{\unhbox\pippobox}\ }
\newcommand\fverbit{\egroup\item[\fbox{\unhbox\pippobox}]}
\newbox\pippobox

\def \lta {\mathrel{\vcenter
     {\hbox{$<$}\nointerlineskip\hbox{$\sim$}}}}
\def \gta {\mathrel{\vcenter
     {\hbox{$>$}\nointerlineskip\hbox{$\sim$}}}}

\newcommand{\bea}{\begin{eqnarray}}
\newcommand{\bdm}{\begin{displaymath}}
\newcommand{\edm}{\end{displaymath}}

\newcommand{\eea}{\end{eqnarray}}

\begin{document}


\begin{center}
{ \Large \bf
Effective Field Theory \\
with a Variable Ultraviolet Cutoff} 
\\
\vspace{1.5cm}
{\Large 
Nikolaos Tetradis 
} 
\\
\vspace{0.5cm}
{\it
Department of Physics, University of Athens,\\
University Campus, Zographou 157 84, Greece
} 
\end{center}
\vspace{3cm}
\abstract{
The properties of strongly gravitating systems suggest that field
theory overcounts the states of a system. Reducing the number of
degrees of freedom, without abandoning the notion of effective field theory,
may be achieved through a connection between the
ultraviolet and infrared cutoffs. 
We provide an implementation of this idea within the Wilsonian approach to the
renormalization group. We derive an
exact flow equation that describes the evolution of the effective action.
We discuss the implications for the existence of infrared fixed points
and the running of couplings.
We also give an alternative derivation in the context of the perturbative
renormalization group.
}

\newpage


{\bf Introduction}:
An effective field theory is characterized by an energy scale $k$ that determines the
region of its applicability. 
The theory is assumed to result from integrating out 
(quantum or thermal) fluctuations of the system with characteristic momenta 
larger than $k$. In this sense, $k$ acts as an infrared cutoff in the process of
integrating out fluctations. In specific applications, $k$ can be identified with
a physical scale, such as the (finite) size of a system or the external momenta in a
certain process. The theory is also assumed to possess an ultraviolet cutoff
$\Lx$. This may viewed simply as a 
technical ingredient in order to regulate the ultraviolet
divergences that often appear when integrating out fluctuations with high momenta. 
However, in the Wilsonian approach to the renormalization group
\cite{wilson}, $\Lx$ corresponds to a microscopic 
physical scale at which the fundamental theory is defined. 
The values of the couplings at this scale determine the fundamental interactions. 
The standard assumption is that $\Lx$ is constant, while the effective 
theory becomes a function of the lower scale $k$.  

The peculiar properties of strongly gravitating systems may indicate that 
we need to abandon the assumption of a fixed $\Lx$ \cite{nelson}. 
One crucial observation is that for an
effective field theory with ultraviolet cutoff $\Lx$
in a box of volume $k^{-3}$ the entropy in general scales $\sim \Lx^3/k^3$. 
On the other hand, the thermodynamics of black holes suggests that the maximum 
entropy must scale with the area and not the volume of the box,
i.e. $S \lta \mpl^2/k^2$, with $\mpl$ the Planck scale
\cite{bekenstein}. 
It is possible that the effective field theory overcounts degrees
of freedom \cite{thooft}.
Reconciling these facts, without abandoning the 
notion of effective field theory, may be achieved through a condition between the 
ultraviolet and infrared cutoffs:
$\Lx \lta \mpl^{2/3} k^{1/3}$.   
A stronger bound is obtained by requiring that the total energy within the 
volume $k^{-3}$ do not lead to a Schwazschild radius for the system 
larger than its size $k^{-1}$.  
If the energy density scales
$\sim \Lx^4$, we must impose that $\Lx^4/k^3 \lta \mpl^2/k$, or 
$\Lx \lta \mpl^{1/2}k^{1/2}$. 

The effective reduction of the ultraviolet 
cutoff $\Lx$ may be attributed to the exclusion of all states that lie within
their Schwazschild radius and cannot be described by an effective local 
quantum field theory \cite{nelson}. 
It is remarkable that the saturation of the stronger bound for an infrared
cutoff of the order of the Hubble scale results in $\Lx \sim 10^{-3}$ eV.
The related energy density is within the observational bounds.  
As a result, the effective reduction in $\Lx$ with diminishing $k$ may be 
relevant for the resolution of the problem of the cosmological constant.

In the scenario that we have in mind, the bulk of the contributions from 
(real of virtual) modes to quantities such as the energy density or the
renormalized couplings is constrained to lie below an energy 
scale that depends on the IR cutoff. Individual very energetic modes may
still exist, but their effect on the above observables is negligible. 
For example, one may limit the temperature $T$ of a system of size $L$ by
demanding that this system does not collapse into a black hole. This
gives $T^2 \lta M_{\rm Pl}/L$.
States with energy much larger than this limit may exist, but their contribution
to the energy density is negligible. Our basic
assumption is that gravity enforces a similar constraint to the contributions
from virtual fluctuations. 
This leads to
limits on the energy density, the quantum-corrected vacuum energy density
(i.e.\ the minimum of the potential)
~\cite{nelson},
the Higgs mass
(i.e.\ the curvature of the potential at its minimum)
~\cite{Banks},
and possibly all couplings
(i.e.\ the derivatives of the potential).
The smallness of the cosmological constant is consistent with such an assumption.

If the IR cutoff is identified with the Hubble scale, it is conceivable that 
the resulting energy density may play the role of the dark energy that drives
the acceleration of the 
cosmological expansion. It must be emphasized, however, that the resulting
equation of state is not consistent with observational data \cite{hsu}.

Our limited understanding of strongly gravitating systems 
prevents the formulation of a precise proposal for the 
mechanism linking the two cutoffs. 
On the other hand, 
links between various energy scales of a theory can often be established on
general grounds. 
For example, it has been pointed out 
in ref. \cite{dvali} that there is a connection between the number $N$ of 
particle species of a theory, the scale $\Lx$ that sets their masses, and
the scale $\mpl$ of the gravitational interactions. 
Arguments based on the physics of black holes imply that 
$N \Lx^2 \lta \mpl^2$ \cite{dvali}. If this bound is saturated, there must be 
a direct link between $\Lx$ and $\mpl$. The variation of one of them must 
be followed by the variation of the other. 

An interesting case, in which
$N$ is very big and leads to a hierarchy between $\Lx$ and $\mpl$, is the
case of compact extra dimensions \cite{arkani}. The number $N$ of Kaluza-Klein 
modes of
the graviton with masses smaller than the scale $\Lx$ 
of the bulk theory is $\sim (\Lx /k)^n$, 
where $1/k$ is the compactification radius and $n$ the
number of extra dimensions. The bound is saturated in this case, with
$\Lx^{2+n}/k^n \sim \mpl^2$. 
A change in $k$ (of possible dynamical origin) would result in the variation 
of either $\Lx$ or $\mpl$. Usually it is assumed that $\Lx$ is the fundamental
scale and $\mpl$ the induced one. However, if $\mpl$ is kept 
constant, $\Lx$ must adjust to the change of $k$.

In this work we study the problem of an effective field theory in 
$d$ dimensions, with a variable
ultraviolet cutoff $\Lx$, within the Wilsonian approach to the renormalization group.  
We do not propose a mechanism underlying the variation of $\Lx$. 
We only make the basic assumption 
that $\Lx$ is a function of the infrared cutoff $k$, so that
variation of $k$ automatically results in the variation of $\Lx$. Only the modes
within the range $[k,\Lx(k)]$ contribute to the renormalization of the couplings.
The theory is defined at a fundamental scale $\Lx_0$ and we assume that we can
set simultaneously $k=\Lx=\Lx_0$. In such a case the theory is determined by the
``tree-level'' action at the scale $\Lx_0$. 

A simple realization of this scenario results if we assume a
relation of the form
\be
\Lx(k)=k^\delta \Lx_0^{1-\delta}
\label{conn} \ee
between the two cutoffs. For $k=\Lx_0$ we also have $\Lx=\Lx_0$. 
In four dimensions with $\Lx_0=\mpl$,
this expression reproduces the relations that result from the saturation of the
two bounds discussed above, for $\delta=1/3$ and 1/2 respectively. 

A clarifying remark is important at this point. Even though we define the theory
at the scale $\Lx_0$, we can view the bare action as being 
the same for all the theories with 
cutoffs $k$ and $\Lx(k)$, for all values of $k$. 
Clearly, this is a consequence of our assumption that
modes above $\Lx(k)$ do not contribute to the renormalization. The picture we
are advocating must be
contrasted to the standard one, in which the bare theory at some $\Lx(k_2)$ would
result from the one at some higher $\Lx(k_1)$ after incorporating the effect of the 
modes with characteristic momenta between $\Lx(k_2)$ and $\Lx(k_1)$. 
According to our assumptions,
these modes decouple because of strong gravity effects.

We emphasize that we do not specify a physical mechanism that is responsible for
the automatic reduction of $\Lx$ as $k$ is lowered. We simply postulate the
relation between $k$ and $\Lx$ and study the consequences for the low-energy 
theory. As we mentioned above, one desirable consequence is that the vacuum
energy is kept automatically within the observational bounds. On the other hand,
the structure of the low-energy theory receives additional modifications, that 
are the focus of our study. 

Throughout this paper we assume that all the dimensionful parameters
are expressed in units of the fundamental scale $\Lx_0$. Equivalently, we can set 
$\Lx_0=1$.

{\bf Intuitive approach}:
We consider a toy model of a scalar field $\phi$ with a $Z_2$ symmetry
$\phi\leftrightarrow -\phi$. 
The effective potential in the one-loop approximation is \cite{jackiw}
\be
U_k^{(1)}(\rho)=V(\rho)+\frac{1}{2(2\pi)^d}\int_k^{\Lx(k)} d^dq
\ln \left(q^2 + V'(\rho)+2\rho V''(\rho) \right),
\label{effone} \ee
where $\rho=\phi^2/2$ and the prime indicates a derivative with respect to $\rho$.
The ``tree-level'' potential $V$ is equal to the effective potential
$U_k$ for $k=\Lx=\Lx_0$. 
In order to obtain a ``renormalization-group improved'' expression,
we take a logarithmic derivative with respect to $k$ and substitute $U_k$ for $V$ in
the resulting integral. This gives
\begin{eqnarray}
\frac{\partial U_k(\rho)}{\partial \ln k} =
&-& 2 v_d
\left[
k^d\ln k^2 -\frac{d \ln \Lx}{d \ln k}\Lx^{d} 
\ln\Lx^2
\right]
\nonumber \\
&-&
2 v_d \left[k^d \ln\left(1+\frac{U'_k+2\rho U''_k}{k^2}\right)
-\frac{d \ln\Lx}{d \ln k}\Lx^{d} 
\ln\left( 1+\frac{U'_k+2\rho U''_k}{\Lx^2} \right)
\right].
\label{impr} \end{eqnarray}
If $k$ and $\Lx$ are related through eq. (\ref{conn}), we have
$d(\ln\Lx)/d (\ln k) =\delta$.

The first term in the r.h.s. describes the variation of the vacuum energy 
as the infrared cutoff is lowered starting from $k=\Lx_0$.
This term is dominated by the ultraviolet cutoff. It results in a 
contribution to the potential which is field independent and scales $\sim \Lx^d$. 
(The integration of this term with respect to $k$ simply reproduces the
integral of eq. (\ref{effone}) with $V=0$.) 
Identifying $\Lx_0$ with the Planck scale and $k$ with the Hubble distance, and
assuming the relation (\ref{conn}) with $\delta=1/2$, brings the 
vacuum energy within the limits imposed by the observations \cite{nelson}.

In this work we are mainly interested in the role of the second term in
the r.h.s. of eq. (\ref{impr}). We would like to understand its implications
for the structure of the low-energy theory. It is
possible to put the approach on more solid grounds 
through a formal derivation of the flow equation for
the potential \cite{pot,review}.

{\bf Exact flow equation}:
We consider a theory of a real scalar field $\chi$, in
$d$ dimensions, with a
$Z_2$-symmetric action $S[\chi]$.
We add to the kinetic term a regulating piece \cite{pot,review}
\be
\Delta S = \frac{1}{2} \int d^d q
\Rh_k(q) \chi^*(q) \chi(q),
\label{twoone} \ee
where $\chi(q)$ are the Fourier modes of the scalar field.
The function $\Rh_k$  is employed in
order to prevent the propagation of modes
with characteristic momenta outside the interval $k^2 \lta q^2 \lta \Lx^2(k)$.
Within this interval the function $\Rh_k(q)$ is
assumed to vanish, so that the respective modes are
unaffected.
For the modes with $q^2 \gta \Lx^2$ or $q^2 \lta k^2$, the effective propagator
$P(q)=(q^2+\Rh_k(q))^{-1}$ is made to vanish by assuming that $\Rh(q)$ diverges. 
We emphasize at this point that the form of $\Rh_k$ is not unique, 
and many alternative choices are possible.

We subsequently introduce sources and
define the generating functional for the connected Green functions
for the action $S + \Delta S$. Through a Legendre
transformation we obtain the
generating functional for the 1PI Green functions
${\tilde \Gamma}_k[\phi]$, where $\phi$ is the expectation value of the
field $\chi$ in the presence of sources.
The use of the modified propagator for the calculation of
${\tilde \Gamma}_k$ results in the effective integration of only the
fluctuations with $k^2 \lta q^2 \lta \Lx^2(k)$. 
Finally, the cutoff-dependent effective action is
obtained by removing the regulating term
\be
\Gamma_k[\phi] = {\tilde \Gamma}_k[\phi] -
\frac{1}{2} \int d^d q
\Rh_k(q) \phi^*(q) \phi(q).
\label{twofour} \ee

For $k =\Lambda=\Lx_0$, $\Gamma_k$ becomes
equal
to the classical action $S$ (no effective integration of modes takes
place). In the opposite
limit $k \rightarrow 0$ we have $\Lx(k) \to 0$ as well, 
so that only low-frequency modes 
contribute to the action. However, for 
$\delta <1$ in eq. (\ref{conn}), $\Lx$ may exceed $k$ by several
orders of magnitude. 

The means for calculating $\Gamma_k$ from $S$ is provided by an exact
flow equation which describes the response of 
$\Gamma_k$ to variations of the infrared cutoff
($t=\ln k$)
\be
\frac{\partial \Gamma_k[\phi]}{\partial t} 
= \frac{1}{2} {\rm Tr} \left[ \left( \Gamma_k^{(2)}[\phi] + \Rh_k \right)^{-1}
\frac{\partial \Rh_k }{\partial \ln k} \right].
 \label{twofive} \ee
Here $\Gamma_k^{(2)}$ is the second functional derivative of the effective
average action with respect to $\phi$. For
real fields it reads in momentum space
$\Gamma_k^{(2)}(q,q') =
{\delta^2 \Gamma_k}/{\delta \phi^*(q) \delta \phi(q')}$,
with
$\phi(-q)=\phi^*(q)$.
The proof of the flow equation proceeds in complete analogy to the case with
constant $\Lx$, and is presented in refs. \cite{pot,review}.
The exact flow equation gives the response of 
$\Gamma_k$ to variations of the scales $k$ and $\Lx(k)$ through a one-loop expression
involving the exact inverse propagator $\Gamma^{(2)}_k$ together with the
cutoffs provided by $\Rh_k$.

{\bf Flow equation for the potential}:
For the solution of eq. (\ref{twofive})
we have to introduce a truncation scheme.
In general, one considers
an expansion of the form
\be
\Gamma_k =
\int d^dx \left[ U_k(\rho)
+ \frac{1}{2} Z_k(\rho) \partial^{\mu} \phi 
\partial_{\mu} \phi
+ ... \right]
\label{twoeight} \ee
where the dots denote terms with higher derivatives of the field.
This turns the flow equation into a system of coupled 
partial differential equations for
$U_k(\rho)$, $Z_k(\rho)$ and the functions appearing in the higher-derivative terms
\cite{pot,review,indices}. For the simplest truncation, which keeps a
general potential $U_k(\rho)$, sets $Z_k=1$ and ignores the higher-derivative terms,
we have  
\begin{eqnarray}
\frac{\partial U_k(\rho) }{\partial \ln k} &=&
\frac{1}{2} \int \frac{d^d q}{(2 \pi)^d}
\,\frac{\partial \Rh_k(q)}{\partial \ln k} 
\,
\frac{1}{q^2
+ \Rh_k(q) + U'_k(\rho) + 2 \rho U''_k(\rho)}
\nonumber \\
&=&2 v_d\, k^d \, \lh^d_0\left(\frac{U'_k(\rho) + 2 \rho U''_k(\rho)}{k^2} \right),
\label{twonine}
 \end{eqnarray}
with
\be
v_d^{-1} = 2^{d+1} \pi^{\frac{d}{2}} \Gamma\left(\frac{d}{2}\right).
\label{threetwo} \ee
Neglecting the effects of wavefunction renormalization by setting $Z_k=1$ means that
we assume that the anomalous dimension of the field vanishes. 
The anomalous dimension can be
taken into account by extending the truncation \cite{review}.

The threshold function 
\be
\lh^d_0(w)=\frac{1}{2} v_d^{-1} k^{-d} \int \frac{d^d q}{(2 \pi)^d}
\,\frac{\partial \Rh_k(q)}{\partial \ln k} 
\,
\frac{1}{q^2+ \Rh_k(q) + k^2 w}
\label{thres} \ee
is a generalization of a similar function defined in the formulation with 
constant $\Lx$ \cite{review}. We also define ``higher'' threshold functions
through
$\lh^d_1=-\partial\lh^d_0(w)/\partial w$ and 
$\lh^{d}_{n+1}=-({1}/{n}){\partial} \lh^d_n(w)/\partial w$ for $n\geq 1$.
The dimensionless ratio $\Rh_k(q)/q^2$ is a function of 
$q^2/k^2$ and $q^2/\Lx^2(k)$. This means that 
the $k$-derivative of $\Rh_k(q)/q^2$ produces terms proportional to its derivatives with
respect to $q^2/k^2$ or $q^2/\Lx^2$. As a result, the integral of eq. (\ref{thres})
is expected to receive contributions mainly from the regions around
$q=k$ and $q=\Lx$.

It is convenient for our investigation to eliminate the field independent
contribution to the potential. This term reproduces the cutoff dependence of the vacuum
energy. Its form is essentially 
given by the first term in the r.h.s. of eq. (\ref{impr}), with
small variations depending on the form of the cutoff function. A detailed discussion
must deal with technical difficulties related to the proper convergence of 
integrals such as (\ref{thres}). The choice of the cutoff function must be
made with care, in order to guarantee this convergence. In order to avoid such
complications and keep the 
presentation simple, we differentiate eq. (\ref{twonine}) with respect to $\rho$ 
and write the flow equation as
\be
\frac{\partial U'_k(\rho) }{\partial \ln k} 
=-2 v_d\, k^{d-2} \, 
\left(3U''_k+2\rho U'''_k \right)\,
\lh^d_1\left(\frac{U'_k(\rho) + 2 \rho U''_k(\rho)}{k^2} \right).
\label{flow} \ee
The integral in the definition of $\lh^d_1$ has better convergence properties
than the one in $\lh^d_0$, so that the choice of a cutoff function is easier. 

{\bf Sharp cutoff}:
Explicit expressions for $\lh^d_n(w)$ can be obtained by choosing specific forms of
the threshold function $\Rh_k(q)$.
For example, we could use
\be
\Rh_k(q) = q^2 
\left[ \frac{1}{\exp \left( - a \left( {q^2}/{\Lx^2(k)} \right)^b \right)
-\exp \left( - a \left( {q^2}/{k^2} \right)^b \right)}-1
\right],
\label{cutoff1} \ee
where $a$, $b$ are
constants that determine the shape of the cutoff.
Another possible choice is
\be
\Rh_k(q) =  q^2 \left[
\frac{1}{\exp \left(  a \left( {q^2}/{k^2} \right)^b \right)-1}
+\frac{1}{\exp \left(  a \left( {q^2}/{\Lx^2(k)} \right)^{-b} \right)-1}
\right].
\label{cutoff2} \ee
For large values of $b$ the momentum integration in the definition of
the threshold functions is dominated by small intervals around
$q=k$ and $q=\Lx$. 
It is straightforward to compute $\lh^d_1$ in the limit $b\to \infty$.
For both the above choices of $\Rh_k$ we find 
\be
\lh^d_1(w)=l^d_1(w)-\frac{d \ln \Lx}{d \ln k} 
\left(\frac{\Lx}{k}\right)^{d-2}\,l^d_1\left(\frac{k^2 w}{\Lx^2} \right),
\label{binf} \ee
where 
\be
l^d_1(w)=\frac{1}{1+w}
\label{sharp} \ee
is the standard form of the threshold function for constant $\Lx$
in the sharp-cutoff limit \cite{review}.
Substituting into eq. (\ref{flow}) we obtain exact agreement with 
eq. (\ref{impr}).

Our derivation provides a formal justification of eq. (\ref{impr}).
This equation results from an exact flow equation in the sharp cutoff 
limit within a truncation of the effective action that keeps a general
potential and a standard derivative term. 

{\bf Existence of fixed points}:
An important issue concerns the possible existence of fixed points of the
theory. These are associated with non-trivial dynamics near second-order
phase transitions. The simplest possibility is that the
fixed-point theory does not depend on $k$, and has 
${\partial \Gamma_k[\phi]}/{\partial t}=0$. It is easy to check that in general
this is impossible to achieve, even for a truncated form of the
effective action. Setting $\partial U'_k/\partial k=0$ in eq. (\ref{flow}),
with the threshold function given by eqs. (\ref{binf}), (\ref{sharp}),
results in an ordinary differential equation for the potential as a function
of $\rho$. However, the explicit dependence on $k$ of the coefficients of
the various terms in this equation implies that its solution will depend
on $k$ as well. The only scale invariant solution corresponds to the
Gaussian fixed point with $U'_k(\rho)=0$.

Another possibility is to search for fixed points for the 
dimensionless potential formed by
multiplying $U_k$ with appropriate powers of $k$.
The flow equation can be reformulated using the 
dimensionless variables
\be 
\rht=k^{2-d} \rho, 
~~~~~~~~~~~~~~~~~~~~~~~
u_k(\rht)=k^{-d}U_k(\rho).
\label{dimles} \ee
It takes the form
\be
\frac{\partial u_k'}{\partial \ln k}=
-2u'_k +(d-2) \rht u''_k -2 v_d \frac{3u''_k+2\rht u'''_k}{1+u'_k}
+2 v_d \frac{d \ln \Lx}{d \ln k}
\left(\frac{\Lx}{k} \right)^{d-2} 
\frac{3u''_k+2\rht u'''_k}{1+\frac{k^2u'_k}{\Lx^2}},
\label{dimflow} \ee
where now the prime denotes a derivative with respect to $\rht$.

In the case of a constant ultraviolet cutoff $\Lx$, the last term in the r.h.s. of the
above equation vanishes. The remaining terms do not depend explicitly on the
running scale $k$. The fixed points correspond to the solutions of the differential
equation that results from setting $\partial u'_k/\partial \ln k =0$. 
The most famous example is the Wilson-Fisher fixed point of the three-dimensional
theory. 

It is clear that the presence of the new term destabilizes most of
the fixed-point solutions. In particular, for $d=3$ 
the last term dominates as $k\to 0$ because $\Lx/k$ diverges.
This implies that the Wilson-Fisher fixed point disappears. 
The same conclusion can be reached for the fixed points of the 
$O(N)$-symmetric scalar theory in three dimensions.  
The only fixed point that survives in all
dimensions is the Gaussian fixed point $u'_k=0$.
Also for $d=2$ and $k$, $\Lx(k)$ related through eq. (\ref{conn}),
fixed points are possible for $k\to 0$.   
They would correspond to solutions of 
\be
-2u'_k  +\frac{1}{4}  
\left( \delta -\frac{1}{1+u'_k} \right)
\left( 3u''_k+2\rht u'''_k \right)=0.
\label{fp2d}
\ee
However, it must be pointed out that the anomalous dimension 
that we neglected is important for the two-dimensional theory \cite{2dmorris},
and must be included in a detailed study of the existence of these fixed points.

{\bf Logarithmic running in four dimensions}:
For $d=4$ and constant $\Lx$, the only known fixed point of the theory is the
Gaussian one. This persists for variable $\Lx(k)$ as well. 
We are interested in studying the behaviour in its vicinity. 
For this, it is convenient to parametrize the potential as 
\be
U_k(\rho)=m^2(k)\,\rho +\frac{1}{4}\lx(k)\,\rho^2+
\frac{1}{6}\sx(k)\,\rho^3
+\frac{1}{24}\nu(k)\,\rho^4 ...
\label{potexp} \ee
Through repeated differentiation of the flow equation we can derive renormalization-group
equations for the couplings appearing in eq. (\ref{potexp}).

At $\rho=0$ the expansion of eq. (\ref{potexp}) results in an infinite system of coupled 
differential equations. They
provide the $\beta$-functions for the generalized couplings of the theory. 
In $d$ dimensions, the first three are
\begin{eqnarray}
\frac{dm^2}{d\ln k}&=&-6v_d k^{d-2} \lx 
\left[ \left( 1+\frac{m^2}{k^2}\right)^{-1} 
-\frac{d \ln \Lx}{d \ln k} 
\left(\frac{\Lx}{k} \right)^{d-2}
\left( 1+\frac{m^2}{\Lx^2}\right)^{-1} \right]
\label{rg1} \\ 
\frac{d\lx}{d\ln k}
&=&+18v_d k^{d-4} \lx^2 
\left[ \left( 1+\frac{m^2}{k^2}\right)^{-2} 
-\frac{d \ln \Lx}{d \ln k} 
\left(\frac{\Lx}{k} \right)^{d-4}
\left( 1+\frac{m^2}{\Lx^2}\right)^{-2} \right]
\nonumber \\
&&-10v_d k^{d-2} \sx 
\left[ \left( 1+\frac{m^2}{k^2}\right)^{-1} 
-\frac{d \ln \Lx}{d \ln k} 
\left(\frac{\Lx}{k} \right)^{d-2}
\left( 1+\frac{m^2}{\Lx^2}\right)^{-1} \right]
\label{rg2} \\
\frac{d\sx}{d\ln k}
&=&-108v_d k^{d-6} \lx^3 
\left[ \left( 1+\frac{m^2}{k^2}\right)^{-3} 
-\frac{d \ln \Lx}{d \ln k} 
\left(\frac{\Lx}{k} \right)^{d-6}
\left( 1+\frac{m^2}{\Lx^2}\right)^{-3} \right]
\nonumber \\
&&+90v_d k^{d-4} \lx\sx 
\left[ \left( 1+\frac{m^2}{k^2}\right)^{-2} 
-\frac{d \ln \Lx}{d \ln k} 
\left(\frac{\Lx}{k} \right)^{d-4}
\left( 1+\frac{m^2}{\Lx^2}\right)^{-2} \right]
\nonumber \\
&&-14v_d k^{d-2} \nu 
\left[ \left( 1+\frac{m^2}{k^2}\right)^{-1} 
-\frac{d \ln \Lx}{d \ln k} 
\left(\frac{\Lx}{k} \right)^{d-2}
\left( 1+\frac{m^2}{\Lx^2}\right)^{-1} \right].
\label{rg3} \end{eqnarray}
These equations can be solved with initial conditions given at $k=\Lx_0$. 
We consider the standard ``renormalizable'' theory with $m^2(\Lx_0)=m^2_w$,
$\lx(\Lx_0)=\lx_w$, $\sx(\Lx_0)=\nu(\Lx_0)=...=0$.

It is apparent that the corrections 
arising from the variation of $\Lx$ for the 
running of couplings higher than the quartic 
are suppressed by some power of $k$. For example, the contributions to the
running of $\sx$ that involve 
non-negative powers of $k$ are multiplied by at least one higher coupling.
As these are assumed to vanish at $k=\Lx_0$, these corrections can be neglected. 
We point out, however, that interesting behaviour, related to first-order
phase transitions \cite{twth}, can be obtained when such corrections become important.

We focus in the following on the mass term $m^2(k)$ and the quartic coupling $\lx(k)$,
assuming vanishing higher couplings. The terms involving $(1+m^2/k^2)$ or
$(1+m^2/\Lx^2)$ to some negative
power are the threshold functions in the sharp cutoff limit. They can be approximated
by 1, as long as the scalar mode remains lighter than the cutoffs. When the mass
term crosses below one of the cutoffs, the corresponding contribution to the
running effectively vanishes.
For $d=4$ and
$k$, $\Lx$ related through eq. (\ref{conn}), the running of $\lx$ is given by 
\be
\frac{d\lx}{d\ln k}=(1-\delta)\frac{9}{16\pi^2}\lx^2.
\label{betal} \ee
This result corresponds to the one-loop $\beta$-function of perturbation
theory for the running of $\lx$.
The higher-loop contributions are related to the terms we omitted in our
truncation (wavefunction renormalization, higher couplings in the potential).
The novel ingredient in eq. (\ref{betal}) is the factor $1-\delta$. The 
evolution of $\lx$ is still logarithmic, but the running scale is not
$k$, but $k^{1-\delta}$. For $\delta$ not very close to zero, this correction
has observable consequences.

In order to obtain the leading contributions to the running of the mass term, we may
omit the logarithmic running of $\lx$ and approximate the threshold functions
by 1. This gives
\be
m^2(k)=\left[ m^2_w-\frac{3\lx_w}{32\pi^2}(1-2\delta)\Lx_0^2\right]
+\frac{3\lx_w}{32 \pi^2}\left(k^2-2\delta \Lx_0k \right).
\label{mass} \ee
The sign of the mass term for $k\to 0$ can be adjusted through the choice of
$m^2_w$ and $\lx_w$. In this way, 
the theory may end up either in the symmetric phase or in
the one with symmetry breaking. 
It is interesting that for $\delta=1/2$ the renormalized mass term $m^2(0)$ is
equal to the ``tree-level'' one $m^2_w$.

{\bf Perturbative derivation}:
The notion of a variable ultraviolet cutoff can be implemented within the 
perturbative renormalization group as well. Let us
consider the possibility that the ultraviolet cutoff $\Lx$ depends on the 
typical energy scale of the process $E$. 
We can also assume that an infrared cutoff $\ell$, that
is a function of $E$ as well, may appear in the bare Green's functions.
The bare couplings are independent of $E$.
We work within renormalized perturbation theory in four-dimensional Minkowski space. 
The Lagrangian of the scalar theory has the form
\begin{eqnarray}
{\cal L}&=& \frac{1}{2}\left(\partial_\mu \phi_0 \right)^2
-\frac{1}{2}m^2_0\phi_0^2-\frac{\lx_0}{8}\phi^4_0
\nonumber \\
&=&\frac{1}{2}\left(\partial_\mu \phi \right)^2
-\frac{1}{2}m^2\phi^2-\frac{\lx}{8}\phi^4
+\frac{1}{2}\dz\left(\partial_\mu \phi \right)^2
-\frac{1}{2}\dm\phi^2-\frac{\dl}{8}\phi^4,
\label{lagrangian} 
\end{eqnarray}
with $\dz=Z-1$, $\dm=m^2_0 Z-m^2$, $\dl=\lx_0Z^2-\lx$.
The renormalized field is $\phi(x)=Z^{-1/2}\phi_0(x)$. 

The renormalized and bare Green's functions are related through 
\be
\langle \Omega | T \phi(x_1) \phi(x_2) ... \phi(x_n) | \Omega \rangle=
Z^{-n/2} \langle \Omega | T \phi_0(x_1) \phi_0(x_2) ... \phi_0(x_n), \Lx_0 | 
\Omega \rangle,
\label{rengreen2} \ee
with $\Lx_0$ some fixed reference value of the ultraviolet cutoff.
The divergences can be absorbed in the wavefunction renormalization $Z$ and
the renormalized coupling $\lx$, that depend on the renormalization scale $M$.
Let us concentrate on the connected Green's functions 
$G^{(n)}(x_1,...x_n)$.  
Under a variation $M\to M +\delta M$ we have $\lx \to \lx+\delta \lx$,
$\phi \to (1 +\delta H)\phi$, with $H=\ln(Z^{-1/2})$.
According to eq. (\ref{rengreen2}), the connected Green's functions satisfy
$G^{(n)}\to (1+n \delta H) G^{(n)}$.
If they are viewed as functions of $M$ and $\lx$, their variation is
$\delta G^{(n)}=({\partial G^{(n)}}/{\partial M})\, \delta M
+({\partial G^{(n)}}/{\partial \lx})\, \delta \lx$.
This gives the Callan-Symanzik equation 
\be
\left[
M\frac{\partial }{\partial M}+\beta \frac{\partial}{\partial \lx}
+n \gamma \right] G^{(n)}(x_1,...,x_n,M,\lx)=0,
\label{cs} \ee
with $\beta=M(\delta \lx/\delta M)$, $\gamma=-M(\delta H/\delta M)$.

It is instructive to study the running of the quartic coupling 
through an explicit calculation. 
At one-loop, the (Fourier transformed) four-point function is  
\be
G^{(4)}=-i \left[ 3\lx +9\lx^2 \left(V(s)+V(t)+V(u) \right) +\dl \right]
\prod_{i=1,...,4} \frac{1}{p^2_i}.
\label{g41l} \ee
We impose as a renormalization condition that the corrections cancel at the
symmetric point $s=t=u=-M^2$. This means that $\dl=-27\lx^2 V(-M^2)$.

For the calculation of $V(-M^2)$ we must use a regularization with an
explicit ultraviolet cutoff. We can use the Pauli-Villars regularization with
scale $\Lx^2$. We also
include a mass-like infrared cutoff $\ell^2$.
We obtain
\be
V(s) = -\frac{1}{32 \pi^2} \ln \frac{\Lx^2}{\ell^2}
-\frac{1}{32 \pi^2} \left[
2+\left(\frac{4\ell^2-s}{|s|} \right)^{1/2} 
\ln\left( 
\frac{\left( 4 \ell^2-s\right)^{1/2}-(|s|)^{1/2}}{\left(4\ell^2-s\right)^{1/2}
+(|s|)^{1/2}}
\right)
\right],
\label{vs} 
\ee
where $\Lx^2 \gg \ell^2$. 
The above expression seems to depend on $\ell$. However, for 
$\Lx^2\gg |s| \gg \ell^2$ it becomes 
\be
V(s)\simeq -\frac{1}{32\pi^2}
\left[2+\ln \frac{\Lx^2}{|s|} \right].
\label{vfin} \ee
The explicit infrared cutoff is irrelevant, as it is 
replaced by the physical energy scale.

We would like to make the ultraviolet cutoff dependent on the energy scale of
the process. 
For this we assume that $\Lx=\Lx(|s|^{1/2})$. 
The renormalization condition requires a
counterterm 
\be
\dl=\frac{9\lx^2}{32\pi^2}
\left[2+\ln \frac{\Lx^2(M)}{M^2} \right].
\label{counte1} \ee
Then,
\be
G^{(4)}=-i \Biggl[ 
3\lx +\frac{9\lx^2}{32 \pi^2}
\Biggl(
\ln \frac{|s|}{M^2}+ \ln \frac{|t|}{M^2} +\ln \frac{|u|}{M^2} 
\nonumber \\
-3\ln \frac{\Lx^2(|s|^{1/2})}{\Lx^2(M)} \Biggr) \Biggr]
\prod_{i=1,...,4} \frac{1}{p^2_i}.
\label{g41la} \ee

The wavefunction renormalization is irrelevant in this order and we can set $\gamma=0$
in eq. (\ref{cs}). The $\beta$-function can be determined from that equation. We find
\be
\beta(\lx)=\frac{9\lx^2}{16 \pi^2}
\left(1-\frac{\partial \ln \Lx(M)}{\partial \ln M} \right).
\label{beta2} \ee
The running coupling, up to second order, is 
\be
\lx=\lx_1+\frac{9\lx_1^2}{16 \pi^2}\left(\ln\frac{M}{M_1}-\ln\frac{\Lx(M)}{\Lx(M_1)} 
\right).
\label{runn2} \ee

In order to make contact with the Wilsonian approach
we choose a reference scale $M_1=\Lx_0$, where $\Lx_0$ should be identified
with $\mpl$ in the scenario of ref. \cite{nelson}. In particular we impose
$\Lambda(\Lambda_0)=\Lambda_0$. The simplest example that realizes this 
assumption has $\Lx(M)=\Lx_0^{1-\delta} M^\delta$ and 
$\partial \ln \Lx/\partial \ln M=\delta$, with constant $\delta$. 
The $\beta$-function of eq. (\ref{beta2}) agrees with the one derived through
the Wilsonian approach and given by eq. (\ref{betal}).
The running coupling becomes
\be
\lx(M)=\lx_w+\frac{9\lx_w^2}{16 \pi^2}\ln\frac{M}{\Lx(M)}
=\lx_w+(1-\delta)\frac{9\lx_w^2}{16 \pi^2}\ln\frac{M}{\Lx_0}.
\label{runn21} \ee
The bare coupling of the Wilsonian approach must be
identified with $\lx_w$ in the above expressions. We have
$\lx(\Lx_0)=\lx_w$. 

We can also consider the bare coupling of the Lagrangian (\ref{lagrangian}).
Up to second order in the perturbative expansion, it is 
\be
\lx_0=\lx+\dl=\lx_w+\frac{9\lx_w^2}{16 \pi^2}.
\label{bareans} \ee
It is constant, as required by the consistency of the discussion.

{\bf Summary and conclusions}:
The main purpose of this paper has been to examine whether it is possible to 
define an effective field theory with an ultraviolet cutoff $\Lx(k)$ that depends on 
the infrared cutoff $k$. This issue cannot be addressed easily within schemes that
employ dimensional regularization, as the notion of a variable ultraviolet cutoff
cannot be implemented. On the other hand, the Wilsonian approach to the renormalization
group provides the ideal framework for the implementation.  
 
We have shown that it is straightforward to transform the notion of integrating 
out high-frequency fluctuations of the fields to that of integrating out
modes within the range of frequencies between the two cutoffs. 
The effective theory is complemented by an exact flow equation that describes
how the generalized couplings vary as both cutoffs are lowered starting from a
common value $k=\Lx=\Lx_0$. Conceptually, the possibility of a variable ultraviolet
cutoff does not pose a problem.

The complications appear in the predictions of the theory. In general, 
the flow equation
receives significant contributions from the regions of momenta around 
both $k$ and $\Lx$.
A first consequence is that some of the fixed points of the theory with fixed $\Lx$
are destabilized by the appearance of explicit dependence on $\Lx(k)$ in the
flow equation. For the scalar theory that we considered, the fixed points in
three dimensions cease to exist, but fixed points in two dimensions
may still appear. The Gaussian fixed point of the scalar theory 
survives in all dimensions, even though the running in its vicinity 
is modified.

Our main interest lies in the four-dimensional theory. Its basic structure can
be maintained, with two possible phases depending on the sign of the renormalized
mass term. The most interesting aspect concerns the logarithmic running of the
quartic coupling. The one-loop $\beta$-function retains its form, but the 
running scale is not $k$, but $k^{1-\delta}$ with $\delta=d\ln\Lx/d\ln k<1$.  

This feature should characterize the running of the dimensionless couplings
of all four-dimensional theories. In particular, we expect the one-loop 
$\beta$-functions of gauge couplings to receive the same modification.
Issues with gauge-invariance in a formulation with explicit cutoffs make the
detailed discussion of this problem very complicated. However, it has been
shown that the 
Wilsonian approach to the renormalization group reproduces the perturbative
loop expansion of the 
$\beta$-function for Yang-Mills theories in the case of constant $\Lx$
\cite{reuter}. A simple dimensional analysis indicates that 
a variable $\Lx(k)$, implemented through cutoff functions similar to the
ones used above, will generate the
same factor $1-\delta$ in the $\beta$-functions of the gauge couplings.

The scale dependence of the couplings, as deduced from 
experimental data, constrains strongly the allowed range of $\delta$. 
Values of $\delta$ larger than 10\% are strongly disfavoured in the case of
the strong coupling $\alpha_s$, as they would modify
its evolution well beyond the limits allowed by
the error bars \cite{pdg}. For the electromagnetic coupling, the limits are
much more stringent, well below 1 \% \cite{strumia}. 
The values $\delta=1/3$ or 1/2, motivated
by considerations of the entropy or energy of strongly gravitating
systems, are in conflict with these bounds. 
It is possible, however, that that such large values of
$\delta$ are relevant only for the gravitational sector, while the 
response of $\Lx$ to changes in $k$ is much weaker in the other
sectors. 

It must be emphasized as well that the possibility to use the running of couplings in
order to constrain the scale dependence of the ultraviolet cutoff is based on
the assumption that the theory retains the same fundamental structure up to 
that cutoff. For example, the
running of the strong coupling near the
GeV scale may become unconventional if the theory of the strong interactions 
remains valid
up to a scale of order $\mpl^{1-\delta}{\rm GeV}^\delta=10^{19(1-\delta)}$ GeV.
If the strong interactions above a scale $\Lx_{eff}$
become part of a more fundamental theory in which the couplings do not run,
the low-energy theory around the GeV scale displays the conventional running because
$\Lx_{eff}$ acts as a constant effective ultraviolet cutoff. 
Modifications would be expected only at energies below 
$10^{19}(\Lx_{eff}/\mpl)^{1/\delta}$ GeV. 
For $\delta=1/2$ and $\Lx_{eff}=10^{10}$ GeV, this value is below the confinement
scale.

{\it Acknowledgments}:
I would like to thank A. Strumia for useful discussions.
This work was supported by the research programs
``Kapodistrias'' of the University of Athens.

\end{document}